\documentclass[prd,preprint,a4paper,showpacs,nofootinbib,superscriptaddress]{revtex4-1}
\usepackage{bm}
\usepackage{amsmath}
\usepackage{latexsym}
\usepackage{amsfonts}
\usepackage{graphicx}
\usepackage{mathrsfs}
\usepackage[utf8]{inputenc}
\usepackage{CJK}
\usepackage{subfigure}
\usepackage{float}
\usepackage{color}
\usepackage{hyperref}
\hypersetup{
	colorlinks=true,
	linkcolor=red,
	citecolor=blue,
}

\begin{document}

\begin{CJK*}{GB}{gbsn}
\title{Full analytical formulas for frequency response of space-based gravitational wave detectors}

\author{Chunyu Zhang}
\email{chunyuzhang@hust.edu.cn}
\affiliation{School of Physics, Huazhong University of Science and Technology,
Wuhan, Hubei 430074, China}

\author{Qing Gao}
\email{Corresponding author. gaoqing1024@swu.edu.cn}
\affiliation{School of Physical Science and Technology, Southwest University, Chongqing 400715, China}

\author{Yungui Gong}
\email{yggong@hust.edu.cn}
\affiliation{School of Physics, Huazhong University of Science and Technology,
Wuhan, Hubei 430074, China}

\author{Bin Wang}
\email{wang\_b@sjtu.edu.cn}
\affiliation{School of Aeronautics and Astronautics, Shanghai Jiao Tong University, Shanghai 200240, China}

\author{Alan J. Weinstein}
\email{ajw@ligo.caltech.edu}
\affiliation{LIGO Laboratory, California Institute of Technology, Pasadena, California 91125, USA }

\author{Chao Zhang}
\email{chao\_zhang@hust.edu.cn}
\affiliation{School of Physics, Huazhong University of Science and Technology,
Wuhan, Hubei 430074, China}

\begin{abstract}
The discovery of gravitational waves, which are ripples of space-time itself, opened a new window to test general relativity,
because it predicts that there are only plus and cross polarizations for gravitational waves. For alternative
theories of gravity, there may be up to six polarizations.
The measurement of the polarization is one of the major scientific goals for future gravitational wave detectors.
To evaluate the capability of the detector, we need to use
the frequency dependent response functions averaged over the source direction and polarization angle.
We derive the full analytical formulas of the averaged response functions for all six possible polarizations and present their asymptotic behaviors based on these analytical formulas.
Compared with the numerical simulation, the full analytical formulas are more efficient and valid for any equal-arm interferometric gravitational
wave detector without optical cavities in the
arms and for a time-delay-interferometry Michelson combination.
\end{abstract}

\maketitle	

\end{CJK*}

\section{Introduction}

The detection of gravitational waves (GWs)
by the Laser Interferometer Gravitational-Wave
Observatory (LIGO) Scientific Collaboration and the Virgo Collaboration
not only announced the dawn of a new era of multimessenger astronomy, but also opened a new window to investigate the property of gravity in the nonlinear and strong field regimes \cite{Abbott:2016blz,Abbott:2016nmj,
	Abbott:2017vtc,Abbott:2017oio,TheLIGOScientific:2017qsa,
	Abbott:2017gyy,LIGOScientific:2018mvr}.
GWs are ripples of space-time itself predicted by
Einstein's theory of general relativity (GR). In GR,
GWs propagate at the speed of light with two transverse polarization states. In alternative theories of
gravity, GWs may have up to six polarizations, and the propagation speed may differ from the speed of light \cite{Eardley:1974nw,Liang:2017ahj,Hou:2017bqj,Gong:2017bru,Gong:2017kim,Gong:2018cgj,Gong:2018ybk,
	Gong:2018vbo,Hou:2018djz,Hou:2018djz}, so
the measurement of polarization states of GWs can be used to test
GR. With the operation of KAGRA \cite{Somiya:2011np,Aso:2013eba}
joining Advanced LIGO \cite{Harry:2010zz,TheLIGOScientific:2014jea} and Advanced Virgo \cite{TheVirgo:2014hva},
the network of ground-based detectors operating in the high frequency band (10-$10^4$ Hz) may measure these polarization states in the next few years. However, there are many gravitational wave (GW) sources,
such as inspiraling galactic binaries, coalescing supermassive black hole binaries, and secondary GWs from ultra-slow-roll inflation \cite{Gong:2017qlj,Yi:2017mxs} emitting low frequency (mHz-1Hz) GWs.
The detection of low frequency GWs will help address numerous astrophysical, cosmological, and theoretical problems.
The proposed space-based GW detectors such as LISA \cite{Danzmann:1997hm,Audley:2017drz},
TianQin \cite{Luo:2015ght}, and TaiJi \cite{Hu:2017mde}
probe GWs in the millihertz frequency band, while DECIGO  \cite{Kawamura:2011zz} operates in the 0.1 to 10 Hz frequency band.
The network of space-based and ground-based GW detectors will start the era of multiband astronomy.
Furthermore, GWs from coalescing supermassive black hole binaries
are continuous in the millihertz band, so a single space-based detector can measure polarizations due to the movement of the detector.

A space-based GW detector such as LISA, TianQin, and TaiJi is designed as an equal-arm interferometric detector without optical cavities in the arms; we call it a Michelson interferometer (MI).
The dominant laser frequency noises experience the same time
delays in the arms and cancel out when the beams are recombined.
Because of the large structure and the movement of the spacecrafts, it is impossible for space-based detectors to maintain the precise equality of the arm lengths.
The time-delay-interferometry (TDI) technique was proposed to solve these problems due to unequal arms \cite{Tinto:1999yr,Armstrong_1999}. There are six different data combinations to cancel the laser frequency noise \cite{Dhurandhar:2002zcl},
and the Michelson combination (MC) is one of them.
The angular response function of the detector represents its capability to capture the gravitational wave coming from a specific direction with certain polarizations. However, the locations of the sources are usually unknown, so instead we use the
angular response function averaging over all sky locations and polarization angles. The integration is not easy to carry out, and it
is time consuming due to the important frequency dependence,
because the arm length is comparable or even larger than the wavelength of in-band GWs.
A lot of effort has been made to obtain the averaged response functions.
For MI, a semianalytical formula which consists of an analytic expression and a definite integral for the tensor mode,
was obtained in the GW frame \cite{Larson:1999we},
and the method was extended to derive semianalytical formulas
which are the sum of an analytic expression and a definite integral
for the other possible polarizations in \cite{Liang:2019pry}. The averaged response functions of all six possible polarizations were
also obtained in \cite{Blaut:2012zz} by Monte Carlo simulation.
For the TDI MC, a similar semianalytical formula,
which is the sum of an analytic expression and a definite integral
for the tensor mode, was given in \cite{Larson:2002xr}, while a full analytical formula
for the tensor mode in the equal arm TDI MC case was successfully derived in \cite{Lu:2019log}.
For all six TDI combinations and all six possible polarizations,
semianalytical formulas, which are the sum of an analytic expression and a definite integral, were derived in \cite{Zhang:2019oet}.
With Monte Carlo simulation,
the averaged response functions of all TDI combinations
for all six possible polarizations
were obtained for LISA in \cite{Tinto:2010hz}.
On the other hand, for a quick calculation of the averaged response function of the tensor mode, the approximate
analytical expression $R\approx \frac{3}{10}[1+0.6(f/f_*)^2]^{-1}$ was widely used for LISA \cite{Cornish:2018dyw}.
The purpose of this paper is to derive full analytical formulas
for the averaged response functions for both MI and equal-arm TDI MC
so that we can evaluate the
capability of space-based interferometric GW detectors efficiently.

The paper is organized as follows. In Sec. \ref{sec2},
we discuss the antenna response functions
for both MI and TDI MC.
In Sec. \ref{sec3}, we work in the detector frame and derive the relationship between the averaged response functions of MI and those of equal-arm TDI MC.
The analytical formulas for their averaged response functions are derived, and their asymptotic behaviors are also analyzed in Sec. \ref{sec4}.
The detail calculations of these formulas are presented in the Appendix.
The paper is concluded in Sec. \ref{sec6}.

\section{Antenna response functions}
\label{sec2}

In terms of the polarization tensor $e^A_{ij}$, the GW signal is
\begin{equation}
\label{hijt}
h_{ij}(t)=\sum_{A} e^A_{ij} h_A(t),
\end{equation} where $h^A(t)$ is the waveform of input GWs,
$A=+,\times,x,y,b,l$ stands for the plus, cross, vector $x$,
vector $y$, breathing, and longitudinal polarizations, respectively.
The GWs detected in the GW observatory are
\begin{equation}
\label{gwst}
s(t)=\sum_A F^A h_A(t),
\end{equation}
where the angular response function $F^A$ for the polarization $A$ is
\begin{equation}
\label{faeq1}
F^A=\sum_{i,j} D^{ij} e^A_{ij},
\end{equation}
and $D^{ij}$ is the detector tensor. For equal arm space-based interferometric detector with a single round-trip light travel
as shown in Fig. \ref{detectorframe},
the detector tensor is
\begin{equation}
D^{ij}=\frac{1}{2}[\hat{u}^i \hat{u}^j T(f,\hat{u}\cdot\hat{\Omega})-\hat{v}^i \hat{v}^j  T(f,\hat{v}\cdot\hat{\Omega})],
\end{equation}
where $\hat{\Omega}$ is the propagating direction of GWs,
$\hat{u}$ and $\hat{v}$ are the unit vectors along the arms of the detector and the normalized antenna transfer function $T(f,\hat{u}\cdot\hat{\Omega})$ is \cite{Estabrook:1975,Schilling:1997id,Cornish:2001qi}
\begin{equation}
\label{transferfunction}
\begin{split}
T(f,\hat{u}\cdot\hat{w})=\frac{1}{2}&
\{\text{sinc}[\frac{f}{2f^*}(1-\hat{u}\cdot\hat{\Omega})]\exp[-i\frac{f}{2f^*}(3+\hat{u}\cdot\hat{\Omega})] \\
& +\text{sinc}[\frac{f}{2f^*}(1+\hat{u}\cdot\hat{\Omega})]\exp[-i\frac{f}{2f^*}(1+\hat{u}\cdot\hat{\Omega})]\},
\end{split}
\end{equation}
here $\text{sinc}(x)=\sin x/x$, $f^*=c/(2\pi L)$ is the transfer frequency of the detector, $c$ is the speed of light,
$L$ is the arm length of the detector, and a monochromatic GW of frequency f is assumed.

\subsection{Michelson Interferometer}

For the space-based equal arm Michelson interferometer without optical cavities,
the response functions are \cite{Liang:2019pry}
\begin{equation}
\label{responsemi}
\begin{split}
F^A_{MI}=&\frac{\sin\left[\frac12u(1-\mu_2)\right]}{2u(1-\mu_2)}e^{-iu(3+\mu_2)/2}\zeta_2^A+\frac{\sin\left[\frac12u(1+\mu_2)\right]}{2u(1+\mu_2)}e^{-iu(1+\mu_2)/2}\zeta_2^A\\
&-\frac{\sin\left[\frac12u(1-\mu_3)\right]}{2u(1-\mu_3)}e^{-iu(3+\mu_3)/2}\zeta_3^A-\frac{\sin\left[\frac12u(1-\mu_3)\right]}{2u(1-\mu_3)}e^{-iu(1+\mu_3)/2}\zeta_3^A,
\end{split}
\end{equation}
where $u= 2\pi f L/c=f/f_*$, $\mu_2\equiv \hat{n}_2\cdot\hat{\Omega}$, $\mu_3\equiv \hat{n}_3\cdot\hat{\Omega}$, $\zeta^A_2\equiv \sum_{i,j}\hat{n}^i_2\hat{n}^j_2 e^A_{ij}$,
$\zeta^A_3\equiv \sum_{i,j}\hat{n}^i_3\hat{n}^j_3 e^A_{ij}$,
$\hat{n}_2$ is the unit vector from $SC_1$ to $SC_2$, and $\hat{n}_3$ is the unit vector from $SC_1$ to $SC_3$.
Using Eq. \eqref{responsemi}, we obtain
the squares of the response functions as
\begin{equation}
\label{squaremi}
\begin{split}
u^2\left|F^A_{MI}\right|^2=&\frac{2\mu_2^2+(1-\mu_2^2)\sin^2u-2\mu_2\sin (u\mu_2)-2\mu_2^2\cos u\cos(u\mu_2)}{4(1-\mu_2^2)^2}(\zeta_2^A)^2\\
&+\frac{2\mu_3^2+(1-\mu_3^2)\sin^2u-2\mu_3\sin(u\mu_3)-2\mu_3^2\cos u\cos(u\mu_3)}{4(1-\mu_3^2)^2}(\zeta_3^A)^2\\
&-\frac{\eta(u,\mu_2,\mu_3)}{2(1-\mu_2^2)(1-\mu_3^2)}\zeta_2^A\zeta_3^A,
\end{split}
\end{equation}
where
\begin{equation}
\label{etaxeq1}
\begin{split}
\eta(u,\mu_2,\mu_3)=&\mu_2\mu_3[\cos u-\cos(u\mu_2)][\cos u-\cos(u\mu_3)]\\
&+[\sin u-\mu_2\sin(u\mu_2)][\sin u-\mu_3\sin(u\mu_3)].
\end{split}
\end{equation}

\subsection{TDI Michelson Combination}
For the TDI equal arm Michelson variable
$X=y_{32,322}-y_{23,233}+y_{31,22}-y_{21,33}+y_{23,2}-y_{32,3}
+y_{21}-y_{31}$,
the response functions are \cite{Zhang:2019oet}
\begin{equation}
\label{responsemc}
\begin{split}
F^A_{MC}=&\frac12\text{sinc}\left[\frac12u(1-\mu_2)\right]e^{-iu(1+\mu_2)/2}\left[e^{-iu}-e^{-3iu}\right]\zeta_2^A\\
&+\frac12\text{sinc}\left[\frac12u(1+\mu_2)\right]e^{iu(1+\mu_2)/2}\left[1-e^{2iu}\right]\zeta_2^A\\
&+\frac12\text{sinc}\left[\frac12u(1-\mu_3)\right]e^{-iu(1+\mu_3)/2}\left[e^{-3iu}-e^{-iu}\right]\zeta_3^A\\
&+\frac12\text{sinc}\left[\frac12u(1+\mu_3)\right]e^{iu(1+\mu_3)/2}\left[e^{2iu}-1\right]\zeta_3^A,
\end{split}
\end{equation}
where $y_{ab}$ is the relative
frequency fluctuations time series measured from the reception at the spacecraft $SC_b$ with transmission from the spacecraft $SC_d$ ($d\neq a$ and $d\neq b$) along the arm $L_a$  \cite{Estabrook:2000ef,Prince:2002hp}, $L_a$ is opposite to $SC_a$, and the index $a=1$, 2, 3 labels the spacecrafts.
For example, $y_{31}$ is the relative frequency fluctuations time series measured from reception at $SC_1$ with transmission from $SC_2$ along $L_3$.
Similarly, the useful notation for delayed data streams are $y_{31,2}=y_{31}(t-L_2)$, $y_{31,23}=y_{31}(t-L_2-L_3)=y_{31,32}$.
The squares of the response functions are
\begin{equation}
\label{squaremc}
\begin{split}
\frac{u^2\left|F^A_{MC}\right|^2}{\sin^2u}
=&\frac{8\mu_2^2+4(1-\mu_2^2)\sin^2u-8\mu_2\sin(u\mu_2)-8\mu_2^2\cos u\cos(u\mu_2)}{(1-\mu_2^2)^2}(\zeta_2^A)^2\\
&+\frac{8\mu_3^2+4(1-\mu_3^2)\sin^2u-8\mu_3\sin(u\mu_3)-8\mu_3^2\cos u\cos(u\mu_3)}{(1-\mu_3^2)^2}(\zeta_3^A)^2\\
&-\frac{8\eta(u,\mu_2,\mu_3)}{(1-\mu_2^2)(1-\mu_3^2)}\zeta_2^A\zeta_3^A.
\end{split}
\end{equation}
Comparing Eq. \eqref{squaremi} with Eq. \eqref{squaremc}, it is easy to see that $\left|F^A_{MC}\right|^2=16\sin^2 u\left|F^A_{MI}\right|^2$, and $\left|F^A_{MC}\right|^2=0$ when $f=nc/(2L)$ with $n$ as an integer number.

\section{The detector coordinate}
\label{sec3}

\begin{figure}[htp]
\centering
\includegraphics[width=0.6\textwidth]{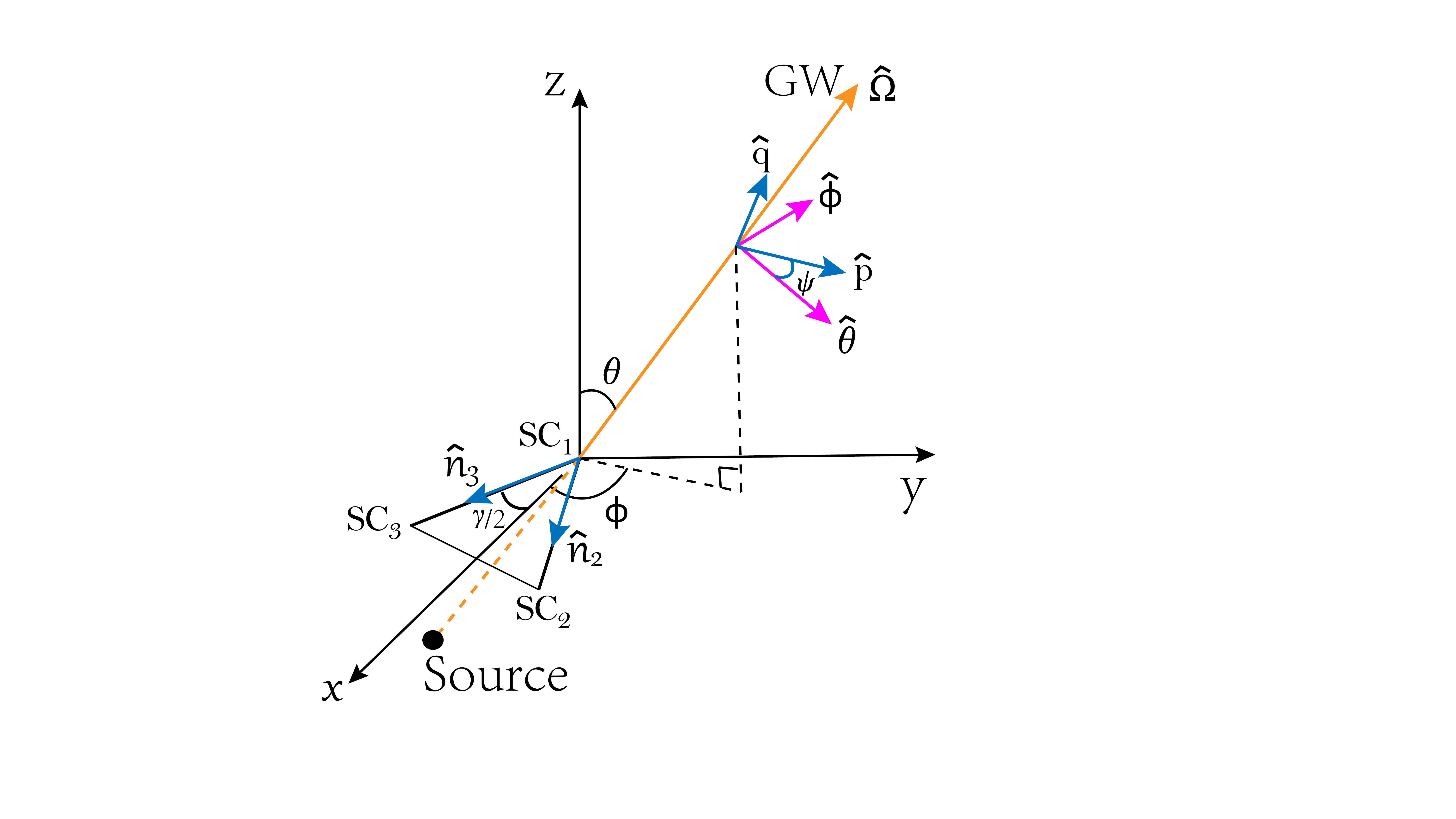}
\caption{The detector frame. GWs propagate along $\hat{\Omega}$, and the laser beam transmits between the spacecrafts.
The unit vector $\hat{n}_2$ is from
the spacecraft $SC_1$ to the spacecraft $SC_2$, and the unit vector $\hat{n}_3$ is from the spacecraft $SC_1$ to the spacecraft $SC_3$.
The arm lengths between the spacecrafts are $L$.}
\label{detectorframe}
\end{figure}

To calculate the averaged response function, we work in the detector coordinate as shown in Fig. \ref{detectorframe}.

\subsection{Basis vectors}

For GWs propagating along the direction $\hat{\Omega}$, we use two
perpendicular unit vectors $\hat{\theta}$ and $\hat{\phi}$
to form an orthonormal coordinate system, such that $\hat{\Omega}=\hat{\theta}\times \hat{\phi}$. In the detector coordinate system, we get
\begin{equation}
\label{basis}
\begin{split}
\hat{\theta}=&(\cos\theta\cos\phi, \cos\theta\sin\phi, -\sin\theta),\\
\hat{\phi}=&(-\sin\phi,\cos\phi,0),\\
\hat{\Omega}=&(\sin\theta\cos\phi,\sin\theta\sin\phi,\cos\theta).
\end{split}	
\end{equation}
The two unit arm vectors are
\begin{equation}
\label{armvec}
\begin{split}
\hat{n}_2=&(\cos\frac\gamma2,\sin\frac\gamma2,0),\\
\hat{n}_3=&(\cos\frac\gamma2,-\sin\frac\gamma2,0),
\end{split}
\end{equation}
where $\gamma$ is the angle between the interferometer's two arms.
Therefore, we get
\begin{equation}
\label{mu12}
\begin{split}
\mu_2=\hat{n}_2\cdot\hat{\Omega}=\sin\theta\cos\phi_-,\\
\mu_3=\hat{n}_3\cdot\hat{\Omega}=\sin\theta\cos\phi_+,
\end{split}
\end{equation}
where $\phi_+=\phi+\gamma/2,\phi_-=\phi-\gamma/2$.
For the convenience of calculation, we introduce $x=\mu_2$ and $y=\mu_3$.

\subsection{The arm scalars}

To account for the rotational degree of freedom around $\hat{\Omega}$, we introduce the polarization angle $\psi$ to form
two new orthonormal vectors $\hat{p}$ and $\hat{q}$,
\begin{equation}
\label{gwvec}
\begin{split}
\hat{p}=&\cos\psi\hat{\theta}+\sin\psi\hat{\phi},\\
\hat{q}=&-\sin\psi\hat{\theta}+\cos\psi\hat{\phi}.
\end{split}
\end{equation}
With the orthonormal vectors $(\hat{p},\hat{q},\hat{\Omega})$, the six polarization tensors are defined as
\begin{equation}
\label{poleij}
\begin{split}
	e^+_{ij}=\hat{p}_i\hat{p}_j-\hat{q}_i\hat{q}_j, \qquad & e^\times_{ij}=\hat{p}_i\hat{q}_j+\hat{q}_i\hat{p}_j, \\
	e^x_{ij}=\hat{p}_i\hat{\Omega}_j+\hat{\Omega}_i\hat{p}_j,
	\qquad & e^y_{ij}=\hat{q}_i\hat{\Omega}_j+\hat{\Omega}_i\hat{q}_j, \\
	e^b_{ij}=\hat{p}_i\hat{p}_j+\hat{q}_i\hat{q}_j, \qquad & e^l_{ij} =\hat{\Omega}_i\hat{\Omega}_j.
\end{split}
\end{equation}
Now we can calculate the variable $\zeta_i^A$.

\subsubsection{The tensor mode}

For the plus and cross modes, we get
\begin{gather}
\zeta^+_2=\sum_{i,j}\hat{n}_2^i\hat{n}_2^je_{ij}^+=(\cos^2\theta\cos^2\phi_--\sin^2\phi_-)\cos(2\psi)-\cos\theta\sin(2\phi_-)\sin(2\psi), \\
\zeta^\times_2=\sum_{i,j}\hat{n}_2^i\hat{n}_2^je_{ij}^\times=-\cos\theta\sin(2\phi_-)\cos(2\psi)-(\cos^2\theta\cos^2\phi_--\sin^2\phi_-)\sin(2\psi),\\
\zeta^+_3=\sum_{i,j}\hat{n}_3^i\hat{n}_3^je_{ij}^+=(\cos^2\theta\cos^2\phi_+-\sin^2\phi_+)\cos(2\psi)-\cos\theta\sin(2\phi_+)\sin(2\psi),\\
\zeta^\times_3=\sum_{i,j}\hat{n}_3^i\hat{n}_3^je_{ij}^\times=-\cos\theta\sin(2\phi_+)\cos(2\psi)-(\cos^2\theta\cos^2\phi_+-\sin^2\phi_+)\sin(2\psi).
\end{gather}
Since we are interested in the average response function for the combined tensor mode $F^T=\sqrt{\left|F^+\right|^2+\left|F^\times\right|^2}$, we use the following arm scalars:
\begin{equation}
\label{armscalart}
\begin{split}
(\zeta^T_2)^2=&(\zeta^+_2)^2+(\zeta^\times_2)^2=(1-x^2)^2,\\
(\zeta^T_3)^2=&(\zeta^+_3)^2+(\zeta^\times_3)^2=(1-y^2)^2,\\
\zeta^T_2\zeta^T_3=&\zeta^+_2\zeta^+_3+\zeta^\times_2\zeta^\times_3=(1-x^2)(1-y^2)-2\sin^2\gamma\cos^2\theta.
\end{split}
\end{equation}

\subsubsection{The vector mode}

For the vector mode, we are interested in the average response function for the combined vector mode $F^V=\sqrt{|F^x|^2+|F^y|^2}$.
Combining Eqs. \eqref{basis}-\eqref{poleij}, it is easy to get
\begin{equation}
\label{armscalarv}
\begin{split}
(\zeta^V_2)^2=&(\zeta^x_2)^2+(\zeta^y_2)^2=4x^2(1-x^2),\\
(\zeta^V_3)^2=&(\zeta^x_3)^2+(\zeta^y_3)^2=4y^2(1-y^2),\\
\zeta^V_2\zeta^V_3=&\zeta^x_2\zeta^x_3+\zeta^y_2\zeta^y_3=4xy(\cos\gamma-xy).
\end{split}
\end{equation}	

\subsubsection{The breathing mode}

For the breathing mode, combining Eqs. \eqref{basis}-\eqref{poleij}, we get
\begin{equation}
\label{armscalarb}
\begin{split}
(\zeta^b_2)^2=&(1-x^2)^2,\\
(\zeta^b_3)^2=&(1-y^2)^2,\\
\zeta^b_2\zeta^b_3=&(1-x^2)(1-y^2).
\end{split}
\end{equation}

\subsubsection{The longitudinal mode}
For the longitudinal mode, combining Eqs. \eqref{basis}-\eqref{poleij}, we get
\begin{equation}
\label{armscalarl}
\begin{split}
(\zeta^l_2)^2=&x^4,\\
(\zeta^l_3)^2=&y^4,\\
\zeta^l_2\zeta^l_3=&x^2y^2.
\end{split}
\end{equation}

\subsection{The averaged response function}

The averaged response (transfer) function is defined as
\begin{equation}
R^A=\frac{1}{8\pi^2}\int_0^\pi\sin\theta d\theta\int_0^{2\pi}d\phi\int_0^{2\pi}d\psi\left|F^A\right|^2.
\end{equation}
Note that $\left|F^A\right|^2$ for the tensor, vector, breathing, and longitudinal modes does not contain the polarization angle $\psi$ and $R^+=R^\times=R^T/2,R^x=R^y=R^V/2$ \cite{Liang:2019pry}; it is unnecessary to integrate over $\psi$ for these modes.
Thus, the averaged response function becomes
\begin{equation}
\label{ra}
\begin{split}
R^A=&\frac{1}{4\pi}\int_0^\pi\sin\theta d\theta\int_0^{2\pi}d\phi\left|F^A\right|^2\\
=&\frac{1}{2\pi}\int_{-1}^1dy\int_{y\cos\gamma-\sin\gamma\sqrt{1-y^2}}^{y\cos\gamma+\sin\gamma\sqrt{1-y^2}}dx\frac{\left|F^A\right|^2}{\sqrt{\sin^2\gamma-x^2-y^2+2xy\cos\gamma}}.
\end{split}
\end{equation}
In the above derivation, we change the integral variables $\theta$ and $\phi$ by
$x=\sin\theta\cos(\phi-\gamma/2)$ and $y=\sin\theta\cos(\phi+\gamma/2)$.
Plugging Eq. \eqref{squaremi} into Eq. \eqref{ra}, we get
\begin{equation}
\label{integralmi}
\begin{split}
u^2R^A_{MI}=&\frac{1}{2\pi}\int_{-1}^1 dy
\int_{y\cos\gamma-\sin\gamma\sqrt{1-y^2}}^{y\cos\gamma+\sin\gamma\sqrt{1-y^2}}\frac{dx}{\sqrt{\sin^2\gamma-x^2-y^2+2xy\cos\gamma}}\\
&\left\{\frac{2x^2+(1-x^2)\sin^2u-2x\sin u\sin(ux)-2x^2\cos u\cos( ux)}{4(1-x^2)^2}(\zeta_2^A)^2\right.\\
&+\frac{2y^2+(1-y^2)\sin^2u-2y\sin u\sin(uy)-2y^2\cos u\cos (uy)}{4(1-y^2)^2}(\zeta_3^A)^2\\
&\left.-\frac{\eta(u,x,y)}{2(1-x^2)(1-y^2)}\zeta_2^A\zeta_3^A\right\},
\end{split}
\end{equation}
where
\begin{equation}
\label{etaxy}
\begin{split}
\eta(u,x,y)=&\sin^2 u-[x\sin(ux)+y\sin(uy)]\sin u+xy\cos[u(x-y)]\\
&+xy[\cos^2 u-\cos u \cos(ux)-\cos u\cos(uy)].
\end{split}
\end{equation}
Because of the relation $\left|F^A_{MC}\right|^2=16\sin^2 u\left|F^A_{MI}\right|^2$,
\begin{equation}
\label{integralmc}
R^A_{MC}=16\sin^2(u)R^A_{MI}.
\end{equation}
Substituting the results in Eqs. \eqref{armscalart},\eqref{armscalarv},\eqref{armscalarb}, and
\eqref{armscalarl} into Eq. \eqref{integralmi}, we can
derive the analytical expressions.
The detailed calculations are presented in the Appendix.

\section{The full analytical formalism}
\label{sec4}
In this section, we present the full analytical formulas of the averaged response functions for interferometric GW detectors without optical cavities. We also give asymptotic behaviors for these averaged response functions.

\subsection{The tensor mode}

The analytical formula of the averaged response function for the combined tensor mode is
\begin{equation}
\label{mit}
\begin{split}
u^2R^T_{MI}=&\frac{3-\cos\gamma}{12}+\frac{-1+\cos\gamma }{u^2}
+2\sin^2\left(\frac{\gamma}{2}\right)\Big[\text{Ci}\left[2u\sin\left(\frac{\gamma}{2}\right)\right]
-\ln\left[\sin\left(\frac{\gamma}{2}\right)\right]\\
&-\text{Ci}(2u)\Big]+\frac{1+\csc^2\left(\frac{\gamma}{2}\right)}{8u^2}
\cos\left[2u\sin\left(\frac{\gamma}{2}\right)\right]+
\sin\left[2u\sin\left(\frac{\gamma}{2}\right)\right]
\left[\frac{-3+\cos\gamma}{32u^3}\right.\\
&\left.+\frac{-21+28\cos\gamma-7\cos(2\gamma)}{32u}\right]\csc^3\left(\frac{\gamma}{2}\right)+\sin(2u)\left[\left(\frac{1}{u}+\frac{2}{u^3}\right)
\sin^2\left(\frac{\gamma}{2}\right)\right.\\
&\left.+\cos^2\left(\frac{\gamma}{2}\right)\Big(2\text{Si}(2u)-\text{Si}\left[2u+2u\sin\left(\frac{\gamma}{2}\right)\right]-\text{Si}\left[2u-2u\sin\left(\frac{\gamma}{2}\right)\right]\Big)\right]
\\
&+\Big[\cos^2\left(\frac{\gamma}{2}\right)\Big(2\text{Ci}(2u)+\ln\left[\cos^2\left(\frac{\gamma}{2}\right)\right]-\text{Ci}\left[2u+2u\sin\left(\frac{\gamma}{2}\right)\right]\\
&-\text{Ci}\left[2u-2u\sin\left(\frac{\gamma}{2}\right)\right]\Big)+\left(\frac{1}{6}-\frac{2}{u^2}\right)
\sin^2\left(\frac{\gamma}{2}\right)\Big]\cos(2u),
\end{split}
\end{equation}
where $\gamma_E$ is the Euler constant, $\text{Si}(u)$
is sine-integral function, and $\text{Ci}(u)$ is cosine-integral function.
Using the analytical expression \eqref{mit},
we plot $R^T_{MI}$ in Fig. \ref{rrfig}, and the plot can be done
in less than a second on a desktop or laptop computer for 500 data points
with $\log(u)$ uniformly distributed from 0.001 to 100.
If we use the semianalytical formula \cite{Larson:1999we,Liang:2019pry},
we need about 6 min to plot the figure as shown with the dotted line in Fig. \ref{rrfig}.
Of course, if we plot more data points, it takes more time.
The analytical and semianalytical formulae give the same results as shown in Fig. \ref{rrfig}.
Therefore, the full analytical expression is much more efficient for the calculation of the transfer function.

In the low frequency limit, $u\ll 1$,
\begin{equation}
\label{lowlim}
\begin{split}
{\rm Si}(2u)\to 2u-\frac{4u^3}{9}+\frac{4u^5}{75}+o(u^7),\\
{\rm Ci}(2u)\to\ln(2u)+\gamma_E-u^2+\frac{u^4}{6}-\frac{2u^6}{135}+o(u^8).
\end{split}
\end{equation}
\begin{equation}
\label{lfmit}
R^T_{MI}\to \frac{2}{5}\sin^2\gamma.
\end{equation}
Note that $\sin(u)/u\to 1$ as $u\to 0$; the terms involving $1/u^2$,
$\cos(u)/u^2$, and $\sin(u)/u^3$ cancel out, and the terms with $\sin(u)/u$ cancel out the first constant term in Eq. \eqref{mit},
so the lowest order is $u^2$ in the right-hand side of Eq. \eqref{mit}.
In the high frequency limit, $u\gg 1$,
\begin{equation}
\label{highlim}
\begin{split}
{\rm Si}(u)\to \frac{\pi}{2}-\frac{\cos(u)}{u}+o(\frac{1}{u^2}),\\
{\rm Ci}(u)\to \frac{\sin(u)}{u}+o(\frac{1}{u^2}).
\end{split}
\end{equation}
\begin{equation}
\label{hfmit}
R^T_{MI} \to R^T_{h1}+R^T_{h2},
\end{equation}
where
\begin{equation}
\label{rthf1}
R^T_{h1}=\left[\frac{3-\cos\gamma}{12}-
2\sin^2\left(\frac{\gamma}{2}\right)\ln\left(\sin\frac{\gamma}{2}\right)\right]\frac{1}{u^2},
\end{equation}
and
\begin{equation}
\label{rthf2}
R^T_{h2}=\left[\frac16\sin^2\left(\frac{\gamma}{2}\right)+\cos^2\left(\frac{\gamma}{2}\right)
\ln\left[\cos^2\left(\frac{\gamma}{2}\right)\right]\right]\frac{\cos(2u)}{u^2}.
\end{equation}

To approximate the averaged response function \eqref{mit}, the following
analytical expression for LISA was widely used \cite{Cornish:2018dyw}:
\begin{equation}
\label{rtapprox1}
R^T\approx \frac{3}{10}\frac{1}{1+0.6(f/f_*)^2}.
\end{equation}
In \cite{Babak:2006uv}, they used the approximation,
\begin{equation}
\label{rtb}
R^T\approx \frac{9}{16(f/f_*)^2}\left[\left(\frac{1}{3}-\frac{2}{(f/f_*)^2}\right)[1+\cos^2(f/f_*)]+\sin^2(f/f_*)+\frac{2\sin(2f/f_*)}{(f/f_*)^3}\right],
\end{equation}
which is the analytical part \eqref{integralb1} of the semianalytical expression for
the averaged response function derived in \cite{Larson:1999we}.
An overall factor of 9/8 is added to the expression so that the low frequency limit is recovered.
It is a coincidence that both the low and high frequency limits are recovered with the same overall factor.
However, for the vector, breathing, and longitudinal modes,
the analytical part in the semianalytical formulas cannot
be used to approximate the full analytical expression because the overall factors for the low and high frequency limits are different.

Based on the expression \eqref{mit} and its low frequency behavior \eqref{lfmit} and  high frequency behavior \eqref{hfmit},
for quick estimation we suggest the following analytical approximation:
\begin{equation}
\label{rtapprox2}
R^T_{MI}\approx \frac{2}{5}\sin^2\gamma \left[1+\frac{2\sin^2\gamma }{5(R^T_{h1}+R^T_{h2})}\right]^{-1}.
\end{equation}
Take $\gamma=\pi/3$, the approximation \eqref{rtapprox2} becomes
\begin{equation}
\label{rtapprox3}
R^T_{MI}\approx \frac{3}{10}\frac{1}{1+(f/f_*)^2/[1.85-0.58\cos(2f/f_*)]}.
\end{equation}
The approximations \eqref{rtapprox1}, \eqref{rtb}, and \eqref{rtapprox3} are shown in Fig. \ref{mifig}. It is interesting that the approximation \eqref{rtb} is more accurate.

\subsection{The vector mode}

The analytical formula of the averaged response functions for the combined vector mode is
\begin{equation}
\label{miv}
\begin{split}
u^2R^V_{MI}=&-4+\frac{4\cos\gamma}{3}+\frac{4-4\cos\gamma }{u^2}
+2\left[\gamma_E-\text{Ci}\left[2u\sin\left(\frac{\gamma}{2}\right)\right]
+\ln\left[2u\sin\left(\frac{\gamma}{2}\right)\right]\right]\\
&-\frac{1+\csc^2\left(\frac{\gamma}{2}\right)}{2u^2}
\cos\left[2u\sin\left(\frac{\gamma}{2}\right)\right]
+\sin\left[2u\sin\left(\frac{\gamma}{2}\right)\right]
\left[\frac{7-8\cos\gamma+\cos(2\gamma)}{8u}\right.\\
&+\left.\frac{3-\cos\gamma}{8u^3}\right]\csc^3\left(\frac{\gamma}{2}\right)
+\left[\left(\frac{4}{u}-\frac{8}{u^3}\right)\sin^2\left(\frac{\gamma}{2}\right)+\text{Si}\left[2u+2u\sin\left(\frac{\gamma}{2}\right)\right]
\right.\\
&\left.
-2\text{Si}(2u)+\text{Si}\left[2u-2u\sin\left(\frac{\gamma}{2}\right)\right]\right]\sin(2u)+\left[\left(\frac{8}{u^2}
-\frac{8}{3}\right)\sin^2\left(\frac{\gamma}{2}\right)-2\text{Ci}(2u)
\right.\\
&\left.+\text{Ci}\left[2u+2u\sin\left(\frac{\gamma}{2}\right)\right]+\text{Ci}\left[2u-2u\sin\left(\frac{\gamma}{2}\right)\right]
-\ln\left[\cos^2\left(\frac{\gamma}{2}\right)\right]\right]\cos(2u).
\end{split}
\end{equation}
Using the analytical expression \eqref{miv}, we plot $R^V_{MI}$ in Figs. \ref{rrfig} and \ref{mifig}.

In the low frequency limit $u\ll 1$,
\begin{equation}
\label{lfmiv}
R^V_{MI}\to \frac{2}{5}\sin^2\gamma.
\end{equation}
In the low frequency limit, the terms involving $1/u^2$,
$\cos(u)/u^2$, and $\sin(u)/u^3$ cancel out, and the terms with $\sin(u)/u$
cancel out the first constant terms in Eq. \eqref{miv}, so $R^V_{MI}$ approaches to a constant.
In the high frequency, limit $u\gg 1$,
\begin{equation}
\label{hfmiv}
\begin{split}
R^V_{MI}\to R^V_h=&
\frac{2}{u^2}\left[\ln\left[\frac{3}{u}+2u\sin\left(\frac{\gamma}{2}\right)\right]
-2+\frac23\cos\gamma+\gamma_E\right]\\
&-\frac{\cos(2u)}{u^2}
\left[\frac83\sin^2\left(\frac{\gamma}{2}\right)
+\ln\left[\cos^2\left(\frac{\gamma}{2}\right)\right]\right].
\end{split}
\end{equation}
For a quick estimation, based on the low and high frequency limits, we may use the approximation,
\begin{equation}
\label{rvapprox1}
R^V_{MI}\approx \frac{2}{5}\sin^2\gamma \left[1+\frac{2\sin^2\gamma}{5 R^V_h}\right]^{-1}.
\end{equation}
Take $\gamma=\pi/3$, the approximation \eqref{rvapprox1} becomes
\begin{equation}
\label{rvapprox3}
R^V_{MI}\approx \frac{3}{10}\frac{1}{1+(f/f_*)^2/[-7.26-1.26\cos(2f/f_*)+6.67\ln(3f_*/f+f/f_*)]}.
\end{equation}
As shown in Fig. \ref{mifig}, the above expression \eqref{rvapprox3} approximates the analytical result \eqref{miv} well.

\subsection{The breathing mode}

The analytical formula of the averaged response function for the breathing mode is
\begin{equation}
\label{mib}
\begin{split}
u^2 R^b_{MI}=&\frac{3-\cos\gamma}{12}+\frac{-1+\cos\gamma}{u^2}
+\sin(2u)\left(\frac{2}{u^3}-\frac{1}{u}\right)\sin ^2\left(\frac{\gamma }{2}\right)\\
&+\sin\left[2u\sin\left(\frac{\gamma }{2}\right)\right]\csc ^3\left(\frac{\gamma }{2}\right) \left[\frac{\cos\gamma -3}{32u^3}+\frac{3-4 \cos\gamma+\cos (2 \gamma )}{32u}\right]\\
&+\cos(2u)\left(\frac{1}{6}-\frac{2}{u^2}\right)\sin^2\left(\frac{\gamma}{2}\right)
+\cos \left[2 u \sin \left(\frac{\gamma }{2}\right)\right]
\frac{1+\csc ^2\left(\frac{\gamma }{2}\right)}{8 u^2}.
\end{split}
\end{equation}
Using the analytical expression \eqref{mib}, we plot $R^b_{MI}$ in Figs. \ref{rrfig} and \ref{mifig}.

In the low frequency limit $u\ll 1$, we get
\begin{equation}
\label{lfmib}
R^b_{MI}\to \frac{1}{15}\sin^2\gamma.
\end{equation}
In the low frequency limit, the terms involving $1/u^2$,
$\cos(u)/u^2$, and $\sin(u)/u^3$ cancel out, and the terms with $\sin(u)/u$
cancel out the first constant term in Eq. \eqref{mib}, so $R^b_{MI}$ approaches to a constant.
In the high frequency limit $u\gg 1$, we get
\begin{equation}
\label{hfmib}
R^b_{MI}\to \frac{3-\cos\gamma+(1-\cos\gamma)\cos(2u)}{12 u^2}.
\end{equation}

For a quick estimation, combining the low and high frequency limits, we may use the approximation,
\begin{equation}
\label{rbapprox1}
R^b_{MI}\approx \frac{1}{15}\sin^2\gamma \left[1+\frac{4u^2 \sin^2\gamma}{5 [3-\cos\gamma+(1-\cos\gamma)\cos(2u)]}\right]^{-1}.
\end{equation}
Take $\gamma=\pi/3$, the approximation \eqref{rbapprox1} becomes
\begin{equation}
\label{rbapprox3}
R^b_{MI}\approx \frac{1}{20}\frac{1}{1+(f/f_*)^2/[4.17+0.83\cos(2f/f_*)]}.
\end{equation}
As shown in Fig. \ref{mifig}, the above expression \eqref{rbapprox3} approximates the analytical result \eqref{mib} well.

\subsection{The longitudinal mode}

\begin{figure}[htp]
\centering
\includegraphics[width=\textwidth]{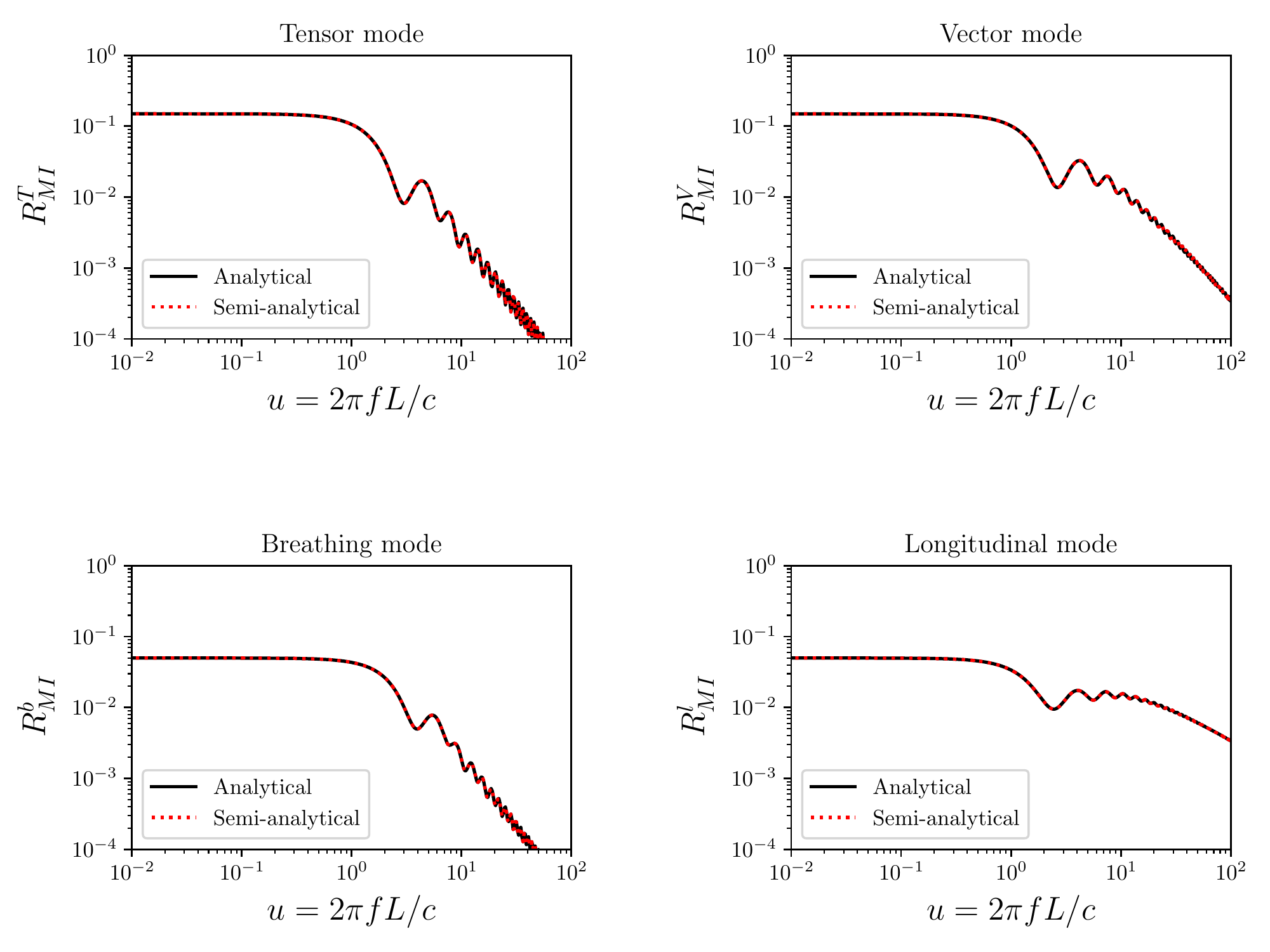}
\caption{The averaged response functions
of different polarizations by using the analytical and semianalytical formulas for interferometric GW detectors without optical cavities in the arms.
We take $\gamma=\pi/3$.
With the analytical expressions, we can plot this figure in less than 3 sec, but it takes more than 40 min to plot this figure if we use the semianalytical formulas derived in \cite{Liang:2019pry}.
The code for the plot and the transfer functions is available at \url{https://github.com/yggong/transfer_function}.}
\label{rrfig}
\end{figure}

\begin{figure}[htp]
\centering
\includegraphics[width=\textwidth]{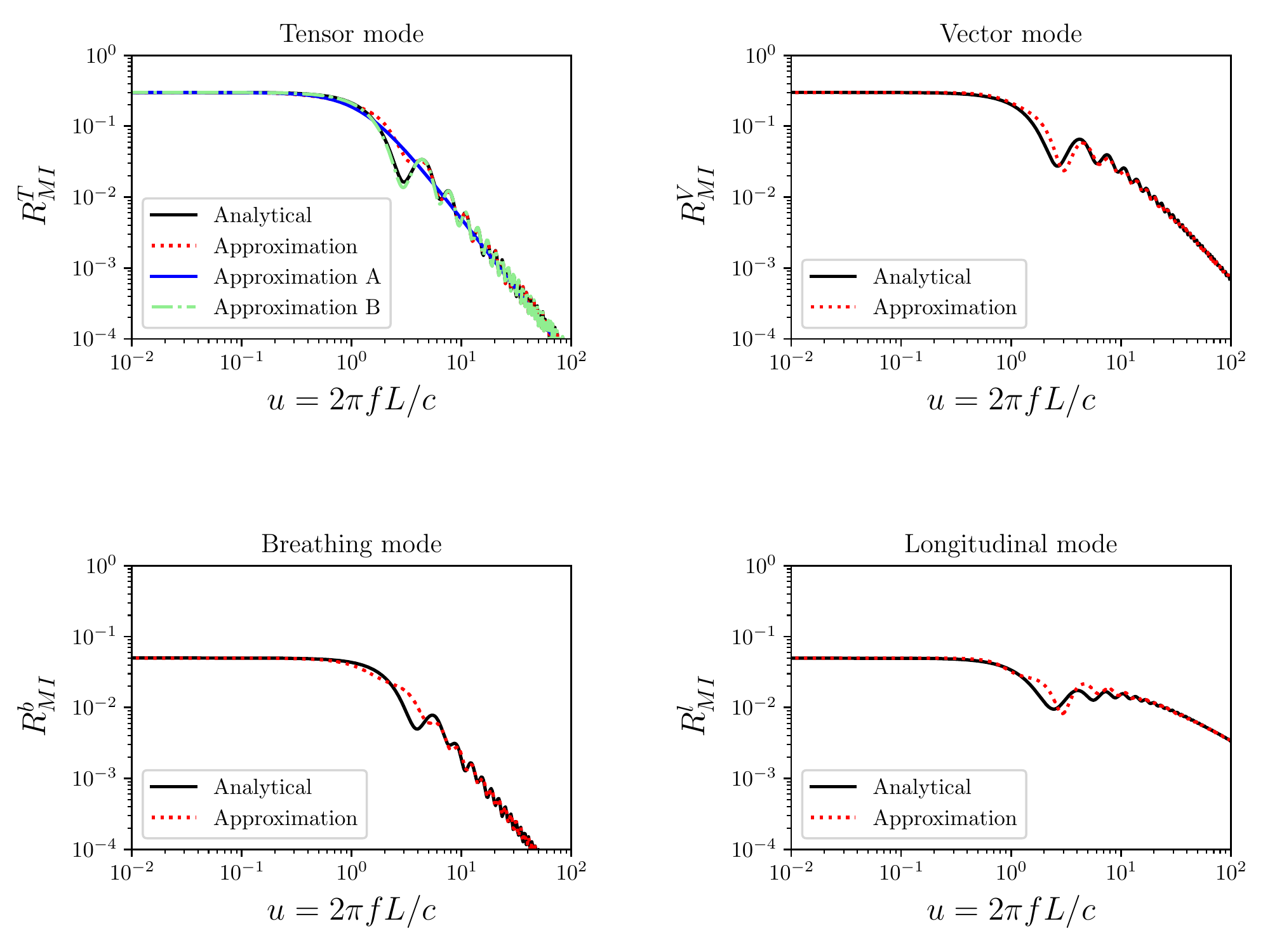}
\caption{The averaged response functions
of different polarizations by using the analytical and approximate formulas for interferometric GW detectors without optical cavities in the arms.
We take $\gamma=\pi/3$. In the upper left panel, the approximation uses the formula \eqref{rtapprox3}, the approximation A uses
the formula \eqref{rtapprox1}, and the approximation B uses the formula \eqref{rtb}.}
\label{mifig}
\end{figure}

The analytical formula of the averaged response functions for the longitudinal mode is
\begin{equation}
\label{mil}
\begin{split}
u^2R^l_{MI}=&\frac{13}{8}-\frac{7\cos\gamma}{12}+\frac{-1+\cos\gamma }{u^2}+\frac{u}{4}\text{Si}(2u)+\left[-\frac78+\frac{\cos\gamma}{4}+\frac{\csc^2\left(\frac{\gamma}{2}\right)}{8}\right]\left[\gamma_E\right.\\
&\left.-\text{Ci}(2u)+\ln(2u)\right]
-\frac14\left[\gamma_E-\text{Ci}\left[2u\sin\left(\frac{\gamma}{2}\right)\right]+\ln\left[2u\sin\left(\frac{\gamma}{2}\right)\right]\right]\csc^2\left(\frac{\gamma}{2}\right)\\
&+\frac{1+\csc^2\left(\frac{\gamma}{2}\right)}{8u^2}\cos\left[2u\sin\left(\frac{\gamma}{2}\right)\right]+\sin\left[2u\sin\left(\frac{\gamma}{2}\right)\right]\csc^3\left(\frac{\gamma}{2}\right)\left[\frac{-3+\cos\gamma}{32u^3}\right.\\
&\left.+\frac{-5+4\cos\gamma+\cos2\gamma}{32u}\right]+\frac12\cos\gamma\cot^2\gamma\left[\sin\left[2u\sin^2\left(\frac{\gamma}{2}\right)\right]\left(\text{Si}\left[2u\sin^2\left(\frac{\gamma}{2}\right)\right]\right.\right.\\
&+\text{Si}\left[2u\sin\left(\frac{\gamma}{2}\right)-2u\sin^2\left(\frac{\gamma}{2}\right)\right]-\text{Si}\left[2u\sin\left(\frac{\gamma}{2}\right)+2u\sin^2\left(\frac{\gamma}{2}\right)\right]\\
&\left.-\text{Si}\left[2u\cos^2\left(\frac{\gamma}{2}\right)\right]\right)+\cos\left[2u\sin^2\left(\frac{\gamma}{2}\right)\right]\Big(\text{Ci}\left[2u\sin^2\left(\frac{\gamma}{2}\right)\right]+\text{Ci}\left[2u\cos^2\left(\frac{\gamma}{2}\right)\right]\\
&\left.-\text{Ci}\left[2u\sin\left(\frac{\gamma}{2}\right)-2u\sin^2\left(\frac{\gamma}{2}\right)\right]-\text{Ci}\left[2u\sin\left(\frac{\gamma}{2}\right)+2u\sin^2\left(\frac{\gamma}{2}\right)\right]\Big)\right]\\
&+\frac{\sec^2\left(\frac{\gamma}{2}\right)}{16}\left[\left(\frac{8}{u^3}-\frac{8}{u}\right)\sin^2\gamma-2\text{Si}\left[2u-2u\sin\left(\frac{\gamma}{2}\right)\right]-2\text{Si}\left[2u+2u\sin\left(\frac{\gamma}{2}\right)\right]\right.\\
&+[4+\cos\gamma-\cos(2\gamma)]\text{Si}(2u)\Big]\sin(2u)+\frac{\sec^2\left(\frac{\gamma}{2}\right)}{16}\left[\frac{10+3\cos\gamma-7\cos(2\gamma)}{3}\right.\\
&+\frac{-4+4\cos2\gamma}{u^2}-[\cos\gamma-\cos(2\gamma)]\gamma_E+[4+\cos\gamma-\cos(2\gamma)]\left[\text{Ci}(2u)-\ln(2u)\right]\\
&-\left.2\text{Ci}\left[2u-2u\sin\left(\frac{\gamma}{2}\right)\right]-2\text{Ci}\left[2u+2u\sin\left(\frac{\gamma}{2}\right)\right]+4\ln\left[2u\cos\left(\frac{\gamma}{2}\right)\right]\right]\cos(2u).
\end{split}
\end{equation}
Using the analytical expression \eqref{mil}, we plot $R^l_{MI}$ in Figs. \ref{rrfig} and \ref{mifig}.

In the low frequency limit, $u\ll 1$,
\begin{equation}
\label{lfmil}
R^l_{MI} \to \frac{1}{15}\sin^2\gamma.
\end{equation}
$R^l_{MI}$ approaches to a constant because
the terms involving $1/u^2$,
$\cos(u)/u^2$, and $\sin(u)/u^3$ cancel out, and
the $\sin(u)/u$ terms
cancel out the first constant terms in Eq. \eqref{mil}
in the low frequency limit.
In the high frequency limit, $u\gg 1$,
\begin{equation}
\label{hfmil}
\begin{split}
R^l_{MI} \to R^l_h=& \frac{\pi}{8u}+\frac{1}{16u^2}-\frac{\ln(2u)}{16 u^2}\left[3+2\csc^2\left(\frac{\gamma}{2}\right)-4\cos\gamma\right.\\
&\left.\quad+\sec^2\left(\frac{\gamma}{2}\right)[\cos\gamma-\cos(2\gamma)]\cos(2u)\right].
\end{split}
\end{equation}
For a quick estimation, based on the low and high frequency limits, we suggest the following approximation:
\begin{equation}
\label{rlapprox2}
R^l_{MI}\approx \frac{1}{15}\sin^2\gamma\left[1+\frac{\sin^2\gamma}{15 R^l_h}\right]^{-1}.
\end{equation}
Take $\gamma=\pi/3$, we get
\begin{equation}
\label{rlapprox3}
R^l_{MI}\approx \frac{1}{20}\frac{1}{1+(f/f_*)^2/[1.25+7.85f/f_*-(11.25+1.67\cos(2f/f_*))\ln(2f/f_*)]}.
\end{equation}
As shown in Fig. \ref{mifig}, the above expression \eqref{rlapprox3} approximates the analytical result \eqref{mil} well.

\begin{figure}[htp]
\centering
\includegraphics[width=\textwidth]{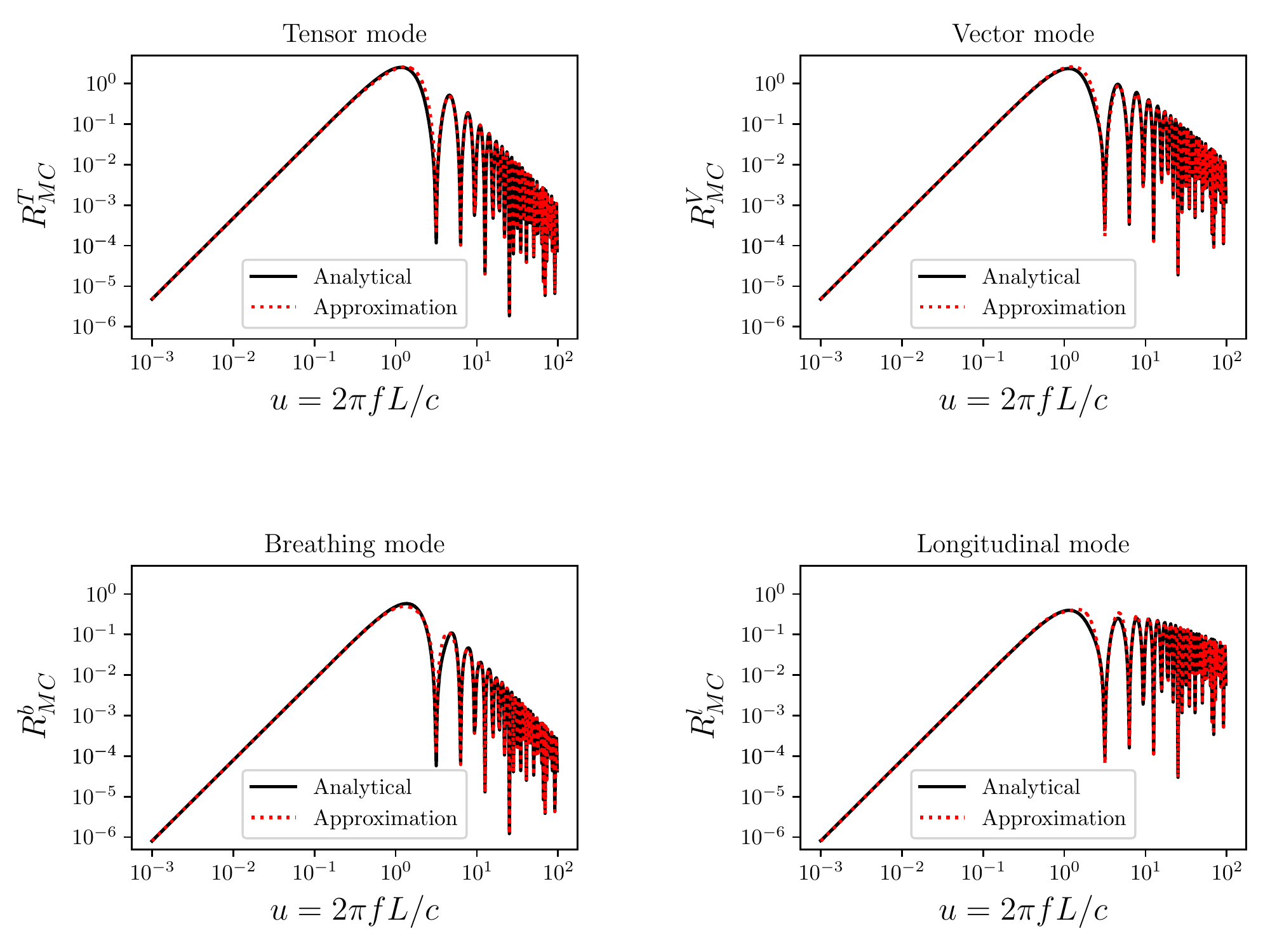}
\caption{The analytical formulas of the averaged response functions for TDI Michelson combination and their approximations with Eqs. \eqref{rtapprox3}, \eqref{rvapprox3}, \eqref{rbapprox3}, and \eqref{rlapprox3} multiplied by $16\sin^2(u)$. We take $\gamma=\pi/3$ in the plots.}
\label{mcfig}
\end{figure}

To obtain the analytical expressions of the transfer functions for
the equal arm Michelson combinations, we use the relation $R^A_{MC}=16\sin^2(u)R^A_{MI}$. The results along with their approximations are shown in Fig. \ref{mcfig}.

\section{Conclusions}
\label{sec6}

For the space-based interferometric GW detectors without optical cavities in the arms, such as LISA, TaiJi, and TianQin,
we derive the full analytical formulas for the frequency dependent response functions for the tensor, vector,
breathing, and longitudinal polarizations. These analytical expressions are consistent with those obtained by Monte Carlo simulation and the semianalytical results we derived before.
With these analytical formulas, the evaluation of the ability of the
detector becomes easier and faster. In particular, the calculation of
the signal to noise ratio for a space-based GW detector becomes
more efficient.
We also find that for space-based interferometric GW detectors, the averaged response functions of the equal arm TDI Michelson combination are just the derived results multiplied by $16\sin^2u$.

With these analytical expressions, the asymptotic behaviors in the low and high frequency limits are apparent. In particular, we can derive analytical expressions for the high frequency behaviors
so that the high frequency behaviors can be easily understood.
For the tensor and breathing modes, the average response functions
fall off as $1/f^2$ at high frequencies, and they also oscillate due
to the factor $\cos(2f/f_*)$.
At high frequencies, $f\gg f_*=c/(2\pi L)$, the averaged response functions decrease as $\ln(f)/f^2$ for the vector mode and $1/f$ for the longitudinal mode.
Even though they also have the term with $\cos(2f/f_*)$, the oscillation is suppressed by a factor of $1/\ln(f)$ or $\ln(f)/f$.
For the vector
mode, the oscillation falls off as $1/f^2$, but the dominant contribution
falls off as $\ln(f)/f^2$. For the longitudinal mode,
the oscillation falls off as $\ln(f)/f^2$, but
the dominant contribution falls off as $1/f$.
For the equal arm TDI Michelson combination, due to the factor
$\sin^2u$, at low frequencies, the averaged response
functions increase as $f^2$. At high frequencies, they
oscillate as $\sin^2(f/f_*)$.

Combining the asymptotic behaviors in the low and high frequency limits, we give simple approximate expressions for
the averaged response functions.
Our approximate expressions to calculate the averaged response functions provide a quick evaluation of the GW detector ability and efficient estimation of the signal to noise ratio. The derived full analytical formulas are useful in space-based GW detection and the
test of theory of gravity.

\begin{acknowledgments}
This research was supported in part by the National Natural Science
Foundation of China under Grants No. 11875136 and 11605061,
the Major Program of the National Natural Science Foundation of China under Grant No. 11690021.
A.J.W. acknowledges funding from the U.S. National Science Foundation, under Cooperative Agreement No. PHY-1764464.
\end{acknowledgments}

\appendix
\section{Detailed calculations for the transfer functions}
\label{appcal}
Because the integration region $x^2+y^2-2xy\cos\gamma\le\sin^2\gamma$ is symmetric for $x$ and $y$, we can interchange $x$ and $y$ in the integrand.
Thus, the first two integrands involving $(\zeta_2^A)^2$ and $(\zeta_3^A)^2$ in Eq. \eqref{integralmi} give the same results.
For the last term with $\zeta^A_2\zeta^A_3$,
the integration of $xy\cos (ux)$ in $\eta(u,x,y)$ is the
same as that of $xy\cos(uy)$,
and the integration of $x\sin(ux)$ in $\eta(u,x,y)$ is the same as that of $y\sin(uy)$. Therefore, we can rewrite $\eta(u,x,y)$ as
\begin{equation}
\label{etaxy1}
\begin{split}
\eta(u,x,y)=&\sin^2 u+xy[\cos^2 u-2\cos u \cos(ux)]-2x\sin(ux)\sin u\\
&+xy\cos[u(x-y)].
\end{split}
\end{equation}
To calculate the last term in Eq. \eqref{integralmi}, we integrate the two lines in Eq. \eqref{etaxy1} separately.

\subsection{The breathing mode}

Substituting Eq. \eqref{armscalarb} into Eq. \eqref{integralmi}, the first two integrations in Eq. \eqref{integralmi} are \cite{Liang:2019pry}
\begin{equation}
\label{integralb1}
\begin{split}
I_{b1}&=\int_{-1}^1dy\int_{y_l}^{y_u}dx\frac{2y^2+(1-y^2)\sin^2u-2y\sin u\sin (uy)-2y^2\cos u\cos (uy)}{4\pi\sqrt{\sin^2\gamma-x^2-y^2+2xy\cos\gamma}}\\
&=\frac{1}{2}\left(\frac{1}{3}-\frac{2}{u^2}\right)(1+\cos^2u)+\frac{1}{2}\sin^2u+\frac{\sin (2u)}{u^3},
\end{split}
\end{equation}
where $y_l=y\cos\gamma-\sin\gamma\sqrt{1-y^2}$ and $y_u=y\cos\gamma+\sin\gamma\sqrt{1-y^2}$. This is the analytical part of the semianalytical formulas obtained in \cite{Liang:2019pry}.

For the integration of the last term in Eq. \eqref{integralmi} isolating the term $xy\cos(ux-uy)$ in $\eta(u,x,y)$, and using the symmetry of $x$ and $y$, we get
\begin{equation}
\label{integralb2}
\begin{split}
I_{b2}=&-\int_{-1}^1dy\int_{y_l}^{y_u}dx \frac{\sin^2u+xy\cos^2u-2xy\cos u\cos (uy)-2y\sin u\sin (uy)}{4\pi\sqrt{\sin^2\gamma-x^2-y^2+2xy\cos\gamma}}\\
=&-\frac14-\frac{\cos\gamma}{12}+\frac{1+2\cos\gamma}{2u^2}+\left(\frac{3-\cos\gamma}{12}-\frac{1-2\cos\gamma}{2u^2}\right)\cos(2u)\\
&-\left(\frac{\cos\gamma}{u^3}+\frac{1-\cos\gamma}{2u}\right)\sin(2u).
\end{split}
\end{equation}
To calculate
the integration of $xy\cos(ux-uy)$, we change the
variables $x$ and $y$ to
$a=(x+y)/2$ and $b=(x-y)/2$. After the change of integration variables, the integration is
\begin{equation}
\label{integralb3}
\begin{split}
I_{b3}=&-\int_{-1}^1dy\int_{y_l}^{y_u}dx \frac{xy\cos (ux-uy)}{4\pi\sqrt{\sin^2\gamma-x^2-y^2+2xy\cos\gamma}}\\
=&-\int_{-\sin\frac{\gamma}{2}}^{\sin\frac{\gamma}{2}}db\int_{-a_l}^{a_l}da \frac{(a^2-b^2)\cos (2ub)}{2\pi\sqrt{\sin^2(\gamma)-4\sin^2(\frac{\gamma}{2})a^2-4\cos^2(\frac{\gamma}{2})b^2}}\\
=&\left(\frac{3-4\cos\gamma+\cos(2\gamma)}{32u}-\frac{3-\cos\gamma}{32u^3}\right)\csc^3\left(\frac{\gamma}{2}\right)\sin\left[2u\sin\left(\frac{\gamma}{2}\right)\right]\\
&+\frac{1+\csc^2\left(\frac{\gamma}{2}\right)}{8u^2}\cos\left[2u\sin\left(\frac{\gamma}{2}\right)\right],
\end{split}
\end{equation}
where $a_{l}=\cos(\gamma/2)\sqrt{1-[b/\sin(\gamma/2)]^2}$.
Add Eqs. \eqref{integralb1}-\eqref{integralb3} together, we get the analytical formula \eqref{mib} for the breathing mode.
Note that
\begin{equation}
\label{bmint1}
I_{b2}+I_{b3}=-\frac{1}{8\pi}\int_0^{2\pi}d\epsilon \int_0^\pi d\theta_1 \sin\theta \eta(u,\tilde{\mu}_1,\tilde{\mu}_2),
\end{equation}
where $\tilde{\mu}_1=\cos\theta_1$, $\tilde{\mu}_2=\cos\gamma\cos\theta_1+\sin\gamma\sin\theta_1\cos\epsilon$ \cite{Liang:2017ahj}.

\subsection{The tensor mode}

Substituting Eq. \eqref{armscalart} into Eq. \eqref{integralmi}, we get the first two integrations in Eq. \eqref{integralmi} as
\begin{equation}
\label{integralt1}
\begin{split}
I_{T1}=I_{b1},
\end{split}
\end{equation}
which gives the analytical part of the semianalytical formulas in \cite{Larson:1999we,Liang:2019pry}.

For the third integration in Eq. \eqref{integralmi}, since $\zeta^T_2\zeta^T_3=(1-x^2)(1-y^2)-2\sin^2\gamma\cos^2\theta$, we integrate the term $(1-x^2)(1-y^2)$ first, and we get
\begin{equation}
\label{integralt2}
\begin{split}
I_{T2}=I_{b2}+I_{b3}.
\end{split}
\end{equation}
To integrate the term $-2\sin^2\gamma\cos^2\theta$, we use the result $\sin\gamma\cos\theta=\sqrt{\sin^2\gamma-x^2-y^2+2xy\cos\gamma}$, so
\begin{equation}
\label{integralt3}
\begin{split}
I_{T_3}=&\frac{1}{4\pi}\int_0^{2\pi}d\epsilon \int_0^\pi d\theta_1 \sin\theta_1 \frac{\sin^2\gamma\sin^2\epsilon}{1-\tilde{\mu_2}^2}\eta(u,\tilde{\mu}_1,\tilde{\mu}_2)\\
=&\int dxdy\frac{\eta(u,x,y)}{2\pi(1-x^2)(1-y^2)}\sqrt{\sin^2\gamma-x^2-y^2+2xy\cos\gamma}\\
=&\frac{-3+4\cos\gamma-\cos(2\gamma)}{4u}\sin\left[2u\sin\left(\frac{\gamma}{2}\right)\right]\csc^3\left(\frac{\gamma}{2}\right)+2\left[\text{Ci}\left[2u\sin\left(\frac{\gamma}{2}\right)\right]\right.\\
&\left.-\text{Ci}(2u)-\ln\left[\sin\left(\frac{\gamma}{2}\right)\right]\right]\sin^2\left(\frac{\gamma}{2}\right)+\left[\frac{2}{u}\sin^2\left(\frac{\gamma}{2}\right)-\left(\text{Si}\left[2u+2u\sin\left(\frac{\gamma}{2}\right)\right]\right.\right.\\
&\left.\left.-2\text{Si}(2u)+\text{Si}\left[2u-2u\sin\left(\frac{\gamma}{2}\right)\right]\right)\cos^2\left(\frac{\gamma}{2}\right)\right]\sin(2u)+\left[\ln\left[\cos^2\left(\frac{\gamma}{2}\right)\right]\right.\\
&\left.+2\text{Ci}(2u)-\text{Ci}\left[2u+2u\sin\left(\frac{\gamma}{2}\right)\right]-\text{Ci}\left[2u-2u\sin\left(\frac{\gamma}{2}\right)\right]\right]\cos^2\left(\frac{\gamma}{2}\right)\cos(2u).
\end{split}
\end{equation}
In the above integration, we follow the method used in the previous section that we separate $\eta(u,x,y)$ into two parts.
Add Eqs. \eqref{integralt1}-\eqref{integralt3} together, we get the analytical formula Eq. \eqref{mit} for the combined tensor mode.

\subsection{The vector mode}

Substituting Eq. \eqref{armscalarv} into Eq. \eqref{integralmi}, the integrations of the first two terms in Eq. \eqref{integralmi} are \cite{Liang:2019pry}
\begin{equation}
\label{integralv1}
\begin{split}
I_{V1}&=\int_{-1}^1dy\int_{y_l}^{y_u}dx\frac{2y^2+(1-y^2)\sin^2u-2y\sin u\sin (uy)-2y^2\cos u\cos (uy)}{\pi(1-y^2)\sqrt{\sin^2\gamma-x^2-y^2+2xy\cos\gamma}}y^2\\
&=-5+\frac{6}{u^2}+2\left[\gamma_E-\text{Ci}(2u)+\ln(2u)\right]+\left(\frac{2}{u}-\frac{4}{u^3}\right)\sin2u+\left(\frac{2}{u^2}-\frac13\right)\cos2u.
\end{split}
\end{equation}
This is the analytical part of the semianalytical formulas obtained in \cite{Liang:2019pry}.
Follow the same procedure, we get the integration of the last
term in Eq. \eqref{integralmi},
\begin{equation}
\label{integralv2}
\begin{split}
I_{V2}=&-\frac{1}{2\pi}\int_0^\pi d\theta_1 \int_0^{2\pi} d\epsilon
 \frac{\cos\theta_1(-\sin\gamma\cos\theta_1\cos\epsilon
 +\cos\gamma\sin\theta_1)\tilde{\mu}_2}{1-
 \tilde{\mu}_2^2} \eta(u,\tilde{\mu}_1,\tilde{\mu}_2)\\
=&-\int_{-1}^1dy\int_{y_l}^{y_u}dx\frac{xy(\cos\gamma-xy)\eta(u,x,y)}{\pi(1-x^2)(1-y^2)\sqrt{\sin^2\gamma-x^2-y^2+2xy\cos\gamma}}\\
=&1+\frac{4\cos\gamma}{3}-\frac{2+4\cos\gamma }{u^2}
+2\left[\text{Ci}(2u)-\text{Ci}\left[2u\sin\left(\frac{\gamma}{2}\right)\right]+\ln\left[\sin\left(\frac{\gamma}{2}\right)\right]\right]\\
&+\frac{1+\csc^2\left(\frac{\gamma}{2}\right)}{2u^2}\cos\left[2u\sin\left(\frac{\gamma}{2}\right)\right]+\sin\left[2u\sin\left( \frac{\gamma}{2}\right)\right]\csc^3\left(\frac{\gamma}{2}\right)\left[\frac{3-\cos\gamma}{8u^3}\right.\\
&+\left.\frac{7-8\cos\gamma+\cos(2\gamma)}{8u}\right]+\left[\text{Si}\left[2u+2u\sin\left(\frac{\gamma}{2}\right)\right]+\text{Si}\left[2u-2u\sin\left(\frac{\gamma}{2}\right)\right]\right.\\
&\left.-2\text{Si}(2u)-\left(\frac{2}{u}-\frac{4}{u^3}\right)\cos\gamma\right]\sin2u+\left[\frac{2-4\cos\gamma}{u^2}+\frac{-3+4\cos\gamma}{3}-2\text{Ci}(2u)\right.\\
&\left.+\text{Ci}\left[2u+2u\sin\left(\frac{\gamma}{2}\right)\right]+\text{Ci}\left[2u-2u\sin\left(\frac{\gamma}{2}\right)\right]-\ln\left[\cos^2\left(\frac{\gamma}{2}\right)\right]\right]\cos2u.
\end{split}
\end{equation}
Add Eqs. \eqref{integralv1}-\eqref{integralv2} together, we get the analytical formula Eq. \eqref{miv} for the vector mode.

\subsection{The longitudinal mode}

Substituting Eq. \eqref{armscalarl} into Eq. \eqref{integralmi}, the first two integrations in Eq. \eqref{integralmi} are \cite{Liang:2019pry}
\begin{equation}
\label{integrall1}
\begin{split}
I_{l1}=&\int_{-1}^1dy\int_{y_l}^{y_u}dx\frac{2y^2+(1-y^2)\sin^2u-2y\sin u\sin (uy)-2y^2\cos u\cos (uy)}{4\pi(1-y^2)^2\sqrt{\sin^2\gamma-x^2-y^2+2xy\cos\gamma}}y^4\\
=&\frac{15}{8}-\frac{3}{2u^2}+\frac98[\text{Ci}(2u)-\ln(2u)-\gamma_E]+\frac{u}{4}\text{Si}(2u)+\left[\frac{1}{u^3}-\frac{1}{u}+\frac{\text{Si}(2u)}{8}\right]\sin(2u)\\
&+\left[\frac{11}{24}-\frac{1}{2u^2}+\frac18\left[\text{Ci}(2u)-\ln(2u)-\gamma_E\right]\right]\cos(2u),
\end{split}
\end{equation}
which gives the analytical part of the semianalytical formulas in \cite{Liang:2019pry}.
Integrating the third term directly
encounters a divergence.
To overcome the divergent problem, we replace the integrand $x^2 y^2$ by $x^2+y^2-2x^2y^2=x^2(1-y^2)+y^2(1-x^2)$
because $x^2 y^2=\cot^{-1}\gamma (x^2+y^2-2x^2y^2)+\csc^2\gamma([\sin^2\gamma-x^2-y^2+2xy\cos(\gamma)]
-2[xy\cos(\gamma)-x^2y^2])-(1-x^2-y^2+x^2y^2)$, and the result is
\begin{equation}
\label{cxintres}
\begin{split}
I_{cx}=&\frac{2}{\pi}\int_{-1}^1 dx\int_{y_l}^{y_u}
\frac{(x^2+y^2-2x^2y^2)\eta(u,x,y)}{(1-x^2)(1-y^2)\sqrt{\sin^2 \gamma-x^2-y^2+2xy\cos\gamma}}dy\\
=&\frac{8 u \sin [u-u \cos (\gamma )]-[\cos (\gamma )-3] \csc ^2\left(\frac{\gamma }{2}\right) \cos \left[2 u \sin \left(\frac{\gamma }{2}\right)\right]}{u^2}\\
&+\frac{\csc ^3\left(\frac{\gamma }{2}\right) \sin \left[2 u \sin \left(\frac{\gamma }{2}\right)\right] \left[\cos (\gamma )+u^2 \cos (2 \gamma )-u^2-3\right]}{2 u^3}\\
&+\frac{4}{u^2} \left[\left(4-2u^2\right) \sin ^2(u)-u \sin(2u)-2u \sin [u(1-\cos\gamma )]\right]\\
&-2 \cos\gamma\left(\frac{1}{3u^3}[16u (u^2-3)\cos ^2(u)+24\sin (2 u)]-\frac{6}{u} \sin (2 u)\right)\\
&+4\sin^2\left(\frac{\gamma}{2}\right)\sin (2 u)\text{Si}(2 u)+4[\sin ^2(u)+\cos\gamma\cos^2(u)]\left[\log (2u)+\gamma_E -\text{Ci}(2 u)\right]\\
&-2 \cos (\gamma ) \left(2 \cos\left[2u \sin^2 \left(\frac{\gamma }{2}\right)\right] \left(\text{Ci}[u(1+ \cos\gamma)]+\text{Ci}[u(1- \cos\gamma)]\right)\right.\\
&\left.+2\sin \left[2u \sin^2 \left(\frac{\gamma }{2}\right)\right] \left(\text{Si}[u(1-\cos\gamma)-\text{Si}[u(1+\cos\gamma)]\right)\right)\\
&+4\cos\gamma \cos\left[2u \sin^2 \left(\frac{\gamma }{2}\right)\right]\left( \text{Ci}\left[2u \sin \left(\frac{\gamma }{2}\right)\left(1-\sin \left(\frac{\gamma }{2}\right)\right)\right]\right.\\
&\left.+\text{Ci}\left[2u \sin \left(\frac{\gamma }{2}\right)\left(1+\sin \left(\frac{\gamma }{2}\right)\right)\right]\right)\\
&+4\cos\gamma \sin\left[2u \sin^2 \left(\frac{\gamma }{2}\right)\right]\left( \text{Si}\left[2u \sin \left(\frac{\gamma }{2}\right)\left(1+\sin \left(\frac{\gamma }{2}\right)\right)\right]\right.\\
&\left.-\text{Si}\left[2u \sin \left(\frac{\gamma }{2}\right)\left(1-\sin \left(\frac{\gamma }{2}\right)\right)\right]\right).
\end{split}
\end{equation}
Combining Eqs. \eqref{integralb2}, \eqref{integralb3}, \eqref{integralt3}, \eqref{integralv2}, and \eqref{cxintres}, we get
\begin{equation}
\label{integrall2}
\begin{split}
I_{l2}=&-\frac{1}{8\pi}\int_0^\pi d\theta_1 \int_0^{2\pi} d\epsilon  \frac{\cos^2\theta_1\tilde{\mu}_2^2}{\sin\theta_1(1-\tilde{\mu}_2^2)}
\eta(u,\tilde{\mu}_1,\tilde{\mu}_2)\\
=&-\int_{-1}^1dy\int_{y_l}^{y_u}dx\frac{x^2y^2\eta(u,x,y)}{4\pi(1-x^2)(1-y^2)\sqrt{\sin^2\gamma-x^2-y^2+2xy\cos\gamma}}\\
=&\frac{-3-7\cos\gamma}{12}+\frac{1+2\cos\gamma }{2u^2}+\left[\frac14+\frac{\cos\gamma}{4}+\frac{\csc^2\left(\frac{\gamma}{2}\right)}{8}\right]\left[\gamma_E-\text{Ci}(2u)+\ln(2u)\right]\\
&-\frac14\left[\gamma_E-\text{Ci}\left[2u\sin\left(\frac{\gamma}{2}\right)\right]+\ln\left[2u\sin\left( \frac{\gamma}{2}\right)\right]\right]\csc^2\left(\frac{\gamma}{2}\right)+\csc^3\left(\frac{\gamma}{2}\right)\left[\frac{-3+\cos\gamma}{32u^3}\right.\\
&\left.+\frac{-5+4\cos\gamma+\cos(2\gamma)}{32u}\right]\sin\left[2u\sin\left(\frac{\gamma}{2}\right)\right]+\frac{1+\csc^2\left(\frac{\gamma}{2}\right)}{8u^2}\cos\left[2u\sin\left(\frac{\gamma}{2}\right)\right]\\
&+\frac12\cos\gamma\cot^2\gamma\left[\sin\left[2u\sin^2\left(\frac{\gamma}{2}\right)\right]\left(\text{Si}\left[2u\sin\left(\frac{\gamma}{2}\right)-2u\sin^2\left(\frac{\gamma}{2}\right)\right]\right.\right.\\
&\left.+\text{Si}\left[2u\sin^2\left(\frac{\gamma}{2}\right)\right]-\text{Si}\left[2u\cos^2\left(\frac{\gamma}{2}\right)\right]-\text{Si}\left[2u\sin\left(\frac{\gamma}{2}\right)+2u\sin^2\left(\frac{\gamma}{2}\right)\right]\right)\\
&+\cos\left[2u\sin^2\left(\frac{\gamma}{2}\right)\right]\Big(\text{Ci}\left[2u\sin^2\left(\frac{\gamma}{2}\right)\right]-\text{Ci}\left[2u\sin\left(\frac{\gamma}{2}\right)-2u\sin^2\left(\frac{\gamma}{2}\right)\right]\\
&\left.+\text{Ci}\left[2u\cos^2\left(\frac{\gamma}{2}\right)\right]-\text{Ci}\left[2u\sin\left(\frac{\gamma}{2}\right)+2u\sin^2\left(\frac{\gamma}{2}\right)\right]
\Big)\right]+\frac{\sec^2\left(\frac{\gamma}{2}\right)}{16}\sin(2u)\Big[\\
&[3-\cos(2\gamma)]\text{Si}(2u)-2\text{Si}\left[2u-2u\sin\left(\frac{\gamma}{2}\right)\right]-2\text{Si}\left[2u+2u\sin\left(\frac{\gamma}{2}\right)\right]\\
&\left.-\left(\frac{4}{u^3}-\frac{4}{u}\right)\left[1+2\cos\gamma+\cos(2\gamma)\right]\right]+\frac{\sec^2\left(\frac{\gamma}{2}\right)}{16}\cos(2u)\left[\frac{4\cos\gamma+4\cos(2\gamma)}{u^2}\right.\\
&+[1+\cos(2\gamma)]\gamma_E-\frac{1+8\cos\gamma+7\cos(2\gamma)}{3}+[3-\cos(2\gamma)][\text{Ci}(2u)-\ln(2u)]\\
&\left.-2\text{Ci}\left[2u-2u\sin\left(\frac{\gamma}{2}\right)\right]-2\text{Ci}\left[2u+2u\sin\left(\frac{\gamma}{2}\right)\right]+2\ln\left[4u^2\cos^2\left(\frac{\gamma}{2}\right)\right]\right].
\end{split}
\end{equation}
Add Eqs. \eqref{integrall1} and \eqref{integrall2} together, we get the analytical formula Eq. \eqref{mil} for the longitudinal mode.

\end{document}